# How to Improve The Accuracy of Equilibrium Molecular Dynamics For Computation of Thermal Conductivity?

Jie Chen,[1] Gang Zhang,[2, *] and Baowen Li[1, 3 &]

[1]Department of Physics and Centre for Computational Science and Engineering, National University of Singapore, Singapore 117546, Singapore

[2]Department of Electronics, Peking University, Beijing 100871, People's Republic of China

[3]NUS Graduate School for Integrative Sciences and Engineering, Singapore 117456, Singapore

## Abstract

Equilibrium molecular dynamics (EMD) simulations through Green-Kubo formula (GKF) have been widely used in the study of thermal conductivity of various materials. However, there exist controversial simulation results which have huge discrepancies with experimental ones in literatures. In this paper, we demonstrate that the fluctuation in calculated thermal conductivity is due to the uncertainty in determination of the truncation time, which is related to the ensemble and size dependent phonon relaxation time. We thus propose a new scheme in the direct integration of heat current autocorrelation function (HCACF) and a nonzero correction in the double-exponential-fitting of HCACF to describe correctly the contribution to thermal conductivity from low frequency phonons. By using crystalline Silicon (Si) and Germanium (Ge) as examples, we demonstrate that our method can give rise to the values of thermal conductivity in an excellent agreement with experimental ones.

[*] Email: zhanggang@pku.edu.cn; Tel: +86 010-62755317
[&] Email: phylibw@nus.edu.sg; Tel: +65 6516 6864; Fax: +65 64641148

Computation of material properties is indispensable in nowadays research. There are standard methods like *ab initio* method and commercial softwares for computation of electronic, optical and magnetic properties. Different simulation approaches such as the continuum models and kinetic theories [1], Monte Carlo calculations [2] and atomistic simulations such as molecular dynamics (MD) calculations [3-7] have been proposed for computation of thermal conductivity.

Among these methods, MD might be the most used one since it can be used to investigate individual materials with realistic crystalline structures. Many processes involved in heat conduction such as boundary scattering, crystal imperfections, and isotope effects can all be included in MD simulations. There exist two formalisms of MD simulations: nonequilibrium molecular dynamics (NEMD) and equilibrium molecular dynamics (EMD). In NEMD, two heat reservoirs with a temperature difference are imposed at the two ends of the materials to mimic the heat source and sink in real experiment. This approach has been used successfully to study the thermal conductivity of 1-Dimensional systems [3], and in design of phononics devices [8]. However, there are two drawbacks in NEMD: (1) in order to suppress the fluctuation of temperature with time, a quite long simulation time is required in NEMD in order to obtain a smooth temperature gradient, thus the system size that NEMD can handle is usually quite small; (2) the value of thermal conductivity calculated depends very much on the types of the heat bath used and the parameters of the heat bath. [9, 10]

In contrast to NEMD, EMD simulation is based on the **Green-Kubo formula** (GKF) derived from the fluctuation-dissipation theorem [11] and linear response theory, which relates thermal conductivity with heat current autocorrelation function (HCACF) [12, 13].

$$\kappa = \frac{1}{3k_B T^2 V} \int_0^\infty dt \left\langle \vec{J}(0) \cdot \vec{J}(t) \right\rangle, \qquad (1)$$

where $k_B$ is the Boltzmann constant, $V$ is the system volume, $T$ is the system temperature, $\vec{J}(t)$ is the heat current, and the angular bracket denotes an ensemble average. In this GKF (1), the entire dynamics of the systems is expressed through the time correlation function, which invokes no assumption about the physical state of the material. With this method, the thermal properties of materials with a large size, even for bulk materials, can be investigated [4-7].

Theoretically, this is a perfect method. Practically, there exists some ambiguity and challenge in implementing the integration of Eq. (1), because in principle the time in Eq. (1) should be infinite long [13] which is computationally infeasible. For any finite time simulation, the accuracy of HCACF calculated from Eq. (1) is limited by the total simulation time, which corresponds to the maximum ensemble average of the correlation function at t=0. HCACF becomes less accurate over time because of the smaller ensemble average one can get from a finite time simulation. Therefore, numerical error (noise) is inevitably introduced into the calculation, and eventually will contaminate HCACF when it decays to a small value. Consequently, HCACF is only reliable up to a finite time (cut-off time). Thus, thermal conductivity can only be calculated from the truncated HCACF which introduce the ambiguity and have caused controversial results in literatures [4-6, 14-16]. For instance, Che *et al.* used a double exponential function to fit HCACF and their calculation result of thermal conductivity of diamond crystal was 60% lower than the experimental value at room temperature [5]. Schelling *et al.* compared exponential fitting with direct integration and argued that the exponential fittings gives rise to an underestimated value of thermal conductivity [16].

In this Letter, we propose a quantitative method to determine the cut-off time of HCACF accurately. Moreover, we provide a correction to the double-exponential-fitting method based on physical argument. We demonstrate by using Silicon and Germanium as examples that our proposed methods can give rise to the values of thermal conductivity of the materials in an excellent agreement with experimental ones.

In our simulations, we restrict ourselves to the high temperature regime where quantum effect can be neglected and classical MD simulation is expected to be highly valid [17]. The Stillinger-Weber (SW) potential [18] is used as it can accurately describe elastic properties and thermal expansion coefficients [18-21]. The heat current is defined as

$$\vec{J}(t) = \sum_i \vec{v}_i \varepsilon_i + \frac{1}{2} \sum_{\substack{ij \\ i \neq j}} \vec{r}_{ij}(\vec{F}_{ij} \cdot \vec{v}_i) + \frac{1}{6} \sum_{\substack{ijk \\ i \neq j \\ j \neq k}} (\vec{r}_{ij} + \vec{r}_{ik})(\vec{F}_{ijk} \cdot \vec{v}_i), \quad (2)$$

where $\vec{F}_{ij}$ and $\vec{F}_{ijk}$ denote the 2-body and 3-body force, respectively. Zero net momentum for the whole system is implicitly required in Eq. (2). Numerically, velocity Verlet algorithm is employed to integrate Newton's equations of motion, and each MD step is set as 0.8fs. A cubic super cell of N×N×N unit cells is used, and periodic boundary conditions are applied to the super cell in all three directions. For each realization, all the atoms are initially placed at their equilibrium positions but have a random velocity according to Gaussian distribution. Canonical ensemble MD with Langevin heat reservoir first runs for $10^5$ steps to equilibrate the whole system at 1000K (Debye temperature of Si is 658K [22]). Then micro-canonical ensemble MD runs for another $2\times10^6$ steps and heat current is collected at each step. After that, thermal conductivity is calculated according to Eq. (1). We refer this method as direct integration, and the corresponding thermal conductivity as "accumulative thermal conductivity $\kappa_a$" in this paper. The final result is averaged over 8 realizations with different initial conditions. We have calculated even more realizations and find that good convergence already can be obtained with 8 realizations.

Figure 1(a) and (d) show the time dependence (0-400ps) of the normalized HCACF for two typical realizations in a 4×4×4 super cell (other realizations are similar). In all the realizations, HCACF has a very rapid decay at the beginning, followed by a long tail which has a much slower decay. This two-stage decaying characteristic of HCACF has been found in the study of various materials [5, 7, 15, 16]. The rapid decay corresponds to the contribution from short wavelength phonons to thermal conductivity, while the slower decay corresponds to the contribution from long wavelength phonons, which is the dominating part in thermal conductivity [5, 15]. Moreover, it is shown in the figure that HCACF decays to approximately zero at a time much shorter than the total simulation time of 400ps. It has also been checked that even for the largest super cell size N=12 considered in our study, the HCACF still can be well relaxed within 400ps, which means the total simulation time of 400ps is adequate for the present study. Therefore, the following part of this paper will be focused on how to improve the accuracy of EMD calculations of thermal conductivity for a given finite total simulation time.

In order to get a quantitative description of the numerical error, we define the relative fluctuation of HCACF as:

$$F(t) = \left| \frac{\sigma(Cor(t))}{E(Cor(t))} \right|, \quad (3)$$

where $\sigma$ and $E$ denote the standard deviation and mean value of HCACF in the time interval (t, t+δ), respectively. In Fig. 1(b) and (e) we plot $F(t)$ for the above-mentioned two typical realizations. δ is chosen as 0.8ps ($10^3$ time steps), and it has been verified that our estimation of $\tau_c$ is insensitive to δ. As shown in Fig. 1(b) and (e), before a critical time, the relative fluctuation of HCACF F(t) maintains a small value (e.g. less than 1) and does not change significantly. This indicates that HCACF is still reliable as there's no large fluctuation in it. After the critical time, F(t) suddenly becomes very large and changes value over time drastically, which is a typical signature of the random noise. This indicates that HCACF has been contaminated and dominated by computational error, and thus is no longer reliable. In our study, we define this critical time (when F(t) becomes larger than one) as the cut-off time, as marked by the green arrow in Fig. 1. We refer this method to estimate $\tau_c$ as "first avalanche" (FA) in this paper. The essence of FA is to only take into account those contributions from HCACF which is before $\tau_c$ and discard the rest part of HCACF as noise. It is worth mentioning that the tiny but non-zero noise can accumulate over time and change the value of accumulative thermal conductivity dramatically (shown in Fig. 1(c) and (f)).

There are some factors which lead to the variation of the cut-off time. For different realizations with the same super cell size, different initial conditions represent different samples of a thermodynamic ensemble, and they are not exactly equivalent to each other in a finite time simulation. This leads to a fluctuation of cut-off time with respect to different realizations as shown in Fig. 1. Moreover, in a larger super cell which contains longer wavelength phonon modes, HCACF decays slower than in a smaller super cell, thus increasing the cut-off time correspondingly. So simply selecting a unique cut-off time for different realizations and different super cell size can lead to suspicious value of thermal conductivity with quite large error bar [16].

Therefore, in order to suppress the fluctuation in the calculation of thermal conductivity, we propose to search the cut-off time and its corresponding accumulative thermal conductivity on a case-by-case basis, for both different realizations and different super cell size. Fig. 2(a) shows the calculation results for thermal conductivity of the crystalline silicon using the direct integration method with case-by-case based FA calculations. Due to the periodic boundary condition, finite size effect exists in the calculated thermal conductivity when simulation domain is small [5, 16]. In our simulations, the thermal conductivity of Si crystal saturates to about 31 W/mK (see Tab. I) when the super cell size $N≥10$ unit cells. This saturated value is in excellent agreement with experimental result of 31.0 W/mK from Ref. [23] at the same temperature. Moreover, a much smaller error bar is obtained in our study compared with those shown in literature [5, 16].

It is worth pointing out that there are other methods to obtain the cut-off time, such as the first dip (FD) method in which $\tau_c$ is the time when the tail of HCACF first decays to zero [4]. For FD method, $\tau_c$ corresponds to the time when $\kappa_a(t)$ first reaches a plateau or a peak. However, we found that the estimation of $\tau_c$ by FD method is not always reliable. When an obvious plateau in $\kappa_a$ can be observed (shown in Fig. 1(c)), FD can make estimation of $\tau_c$ quite close to that estimated by FA. Consequently, both two methods can give estimation of thermal conductivity quite close to each other. In this case, the mean value of the normalized HCACF fluctuates around zero in a relatively short time (see insets in Fig. 1(b)). As a result, the noise doesn't accumulate over time but gives rise to a plateau in $\kappa_a$ with small fluctuations. However, for the case without the presence of obvious plateau in $\kappa_a$ (shown in Fig. 1(f)), FD estimates $\tau_c$ (according to the peak in $\kappa_a$) to be about 52ps (marked by black arrow), while F(t) shows that HCACF has been already contaminated by noise after about 15ps. In this case, FD overestimates the thermal conductivity by falsely taking into account the contribution from the positive noise (see insets in Fig. 1(e)). This indeed can become a very serious problem when the noise remains positive over a very long time in some realizations (not shown here). Therefore, in this Letter, the cut-off time in all the realizations are estimated by FA.

We have demonstrated that direct integration with FA on a case-by-case basis can be quite efficient to make a successful prediction of thermal conductivity and suppress the error bar. However, EMD simulation with direct integration becomes a formidable task when dealing with large systems with a long mean free path. Several alternatives have been proposed by making use of certain statistical properties of HCACF [4-6, 14-16]. The single exponential function was first used to fit HCACF, which is then integrated analytically to get the thermal conductivity [4]. Due to the two-stage decaying characteristic of HCACF mentioned above, the single exponential fitting is unable to capture the statistics of HCACF fully, leading to an underestimation of thermal conductivity [16]. A similar fitting approach in frequency domain based on the single exponential decay of HCACF was proposed by Volz and Chen [6], but it was also found to have the same problem of underestimating thermal conductivity [16]. Che *et al.* developed a double exponential function to fit HCACF. However their calculation result of thermal conductivity of diamond crystal was also 60% lower than the experimental value at room temperature [5].

In the following part, we demonstrate the availability of a non-zero correction in the double exponential fitting approach. In our method, HCACF is fitted according to the following function

$$Cor(t)/Cor(0) = A_1 e^{-t/\tau_1} + A_2 e^{-t/\tau_2} + Y_0, \qquad (4)$$

in which $A_1$, $A_2$, $\tau_1$, $\tau_2$, and $Y_0$ are fitting parameters. The decay of HCACF in bulk material will be exponential due to the macroscopic law of relaxation and Onsager's postulate for microscopic thermal fluctuation [5]. It has been reported that the relaxation times generally decreases with increasing phonon frequency. [24] In bulk Silicon, the relaxation times of acoustic phonon (both longitudinal and transverse) are from one to hundreds ps, with the phonon frequency range is from 0 to 13 THz. However, for optical phonons (both longitudinal and transverse) with frequency >13 THz, the relaxation times are from 1 to 10 ps, much shorter than those for acoustic phonons. [24] Thus in HCACF, the initial fast decay is due to the high frequency optical modes, while the slow decay corresponds to the low frequency acoustic modes. The time constant $\tau_1$ and $\tau_2$ correspond to the relaxation time of *short wavelength and long wavelength phonons*, respectively. Levenberg-Marquardt algorithm is used in the nonlinear least-square fitting. The difference between Eq. (4) and the one used in Ref. [5] is the constant $Y_0$. In our calculations, we first search the cut-off time $\tau_c$ based on FA method, and then HCACF before $\tau_c$ is fitted according to Eq. (4). Finally a finite time of integration up to $\tau_c$ is performed and the thermal conductivity is obtained by the following equations

$$\kappa_C(t) = \frac{Cor(0)}{3k_B T^2 V}(A_1\tau_1 + A_2\tau_2) \qquad (5)$$

$$\kappa_F(t) = \frac{Cor(0)}{3k_B T^2 V}(A_1\tau_1 + A_2\tau_2 + Y_0\tau_c), \qquad (6)$$

in which $\kappa_c$ and $\kappa_F$ means excluding and including the contribution from $Y_0$, respectively.

Fig. 2(a) shows the calculation results based on double-exponential-fitting without and with $Y_0$. In both two cases, thermal conductivity saturates to a constant value with the increase of super cell size. Without $Y_0$, the saturated value for $\kappa_c$ is about 16 W/mK (see Tab. I), which is only about half of the experimental value. This underestimation of thermal conductivity is quite similar to what was found in Ref. [5, 6, 16]. With $Y_0$ correction, which is typically on the order of $10^{-3}$ for most of the realizations, the saturated value for $\kappa_F$ is about 31 W/mK, in good agreement with experimental value [23]. As shown in Table I, double-exponential-fitting with $Y_0$ actually reproduces the same result as that calculated from direct integration, which indicates that this tiny but nonzero term $Y_0$ is physical rather than just a numerical error.

Fig. 3 shows the raw data and the corresponding fitted curve according to double-exponential-fitting (Eq. 4) of the normalized HCACF before the cut-off time for the same two realizations shown in Fig. 1. The fitted curve can well fit the slow decay region of HCACF, which makes the dominating contribution to the thermal conductivity. The insets show the long time region near cut-off time (marked by the green arrow). At this region, although the raw data of HCACF can reach zero due to the fluctuation of noise, its average contribution (envelope) over time is not zero due to the finite cut-off time, and corresponds to the tiny term $Y_0$. *This term is the contribution from long wavelength phonons which have a longer relaxation time than the cut-off time.* In bulk Silicon, for acoustic phonons with frequency less than 2THz, their relaxation times are longer than about 50 ps (generally $>\tau_c$). Moreover, it has been demonstrated that the phonons those have frequency less than 2THz contribute ~50% to the thermal conductivity of bulk silicon at 1000K. [24] This is in good agreement with our calculations that excluding $Y_0$, the calculated thermal conductivity is only about half of the experimental value. This is the physical origin of the tiny but nonzero $Y_0$.

In order to test the validity of our correction to double-exponential-fitting, we have also calculated the thermal conductivity of crystalline Germanium (Debye temperature of Ge is 372K [22]) at 1000K. A finite time of cut-off is used in all the calculations according to first avalanche. As shown in Fig. 2(b), the saturated values of thermal conductivity calculated from direct integration and double exponential fitting with $Y_0$ are both about 17 W/mK, which is in good agreement with experimental value of 17.1 W/mK from Ref. [23] at the same temperature. However, excluding $Y_0$ in the double exponential fitting predicts an underestimated value of about 10 W/mK.

Now we turn to discuss the possible factors that may influence the accuracy of our proposed double-exponential-fitting method. As shown in Fig. 3, our method can fit the slow decay region of HCACF (relates to τ2) very well, but cause deviation from the raw data in the fast decay region (relates to τ1). However, the time constant τ2 for long wavelength phonons is more than 100 times larger than τ1 for short wavelength phonons. As a result, the contributions from short wavelength phonons only account for less than 2% of the overall thermal conductivity. Therefore, this deviation of fitted curve from raw data has little effect on the accuracy of the calculated thermal conductivity.

The total simulation time is the crucial factor that limits the accuracy of our proposed method. An insufficient total simulation time will lead to the insufficient ensemble average in the calculation of HCACF. Moreover, HCACF may not be well relaxed if the intrinsic relaxation time is larger than the total simulation time. This can cause uncertainty in determining the time constant τ1 and τ2 among different realizations, which eventually may compromise the accuracy of calculated thermal conductivity based on double-exponential-fitting. However, as we have already demonstrated in the first part of this study, the total simulation time in our study is adequate enough to ensure the accuracy and relaxation of HCACF. As a result, for all the super cell size considered in this study, we obtain less than 9% deviation in τ2 for long wavelength phonons, and less than 2% deviation in τ1 for short wavelength phonons among different realizations.

Moreover, here we only carry out calculation at ambient temperature which is higher than Debye temperature because classical molecular dynamics is only valid in this regime, thus we can solely test the validity of our proposed method without considering quantum effect. To calculate thermal conductivity at ambient temperature which is lower than Debye temperature, quantum correction [17] to classical MD calculations must be considered. In addition, in the case where the mean free path is much longer than the box size, the HFACS tails cannot be calculated. Thus a larger system size and longer simulation time are needed in order to get well relaxed HCACF. It deserves further investigation on how to combine quantum correction and our proposed approach to calculate thermal conductivity for larger mean free path cases.

In summary, we have examined different implementations of Green-Kubo formula in EMD simulations. Due to the finite number of ensemble average, HCACF is only reliable up to a cut-off time and thus the thermal conductivity can only be calculated from the truncated HCACF. We have proposed an efficient *quantitative* method (*first avalanche*) to accurately estimate the cut-off time. Using the cut-off time, direct integration method can make a successful computation of thermal conductivity of crystalline silicon and germanium. In addition, we have demonstrated that because of the finite cut-off time, a small nonzero correction term can significantly improve the accuracy of EMD calculations based on the double-exponential-fitting of HCACF. Excluding this term in the calculation gives rise to an underestimated value of thermal conductivity due to the partial exclusion of contribution from low-frequency phonons, while including it one can make a correct prediction in good agreement with experimental value. Since the two-stage exponential decaying characteristic of HCACF has been found in various materials and has profound underlying physical mechanism, our method is quite general and can have wide applications in accurate thermal conductivity estimations of different materials and systems.

**Acknowledgement**

JC is indebted to Prof. Jian-Sheng Wang for helpful discussions about Green-Kubo formula and EMD simulations. This work is supported in part by an ARF grant, R-144-000-203-112, from the Ministry of Education of the Republic of Singapore, and Grant R-144-000-222-646 from National University of Singapore.

FIG. 1. (Color online) Time dependence of normalized heat current autocorrelation function Cor(t)/Cor(0) (red line), relative fluctuation of Cor(t) (violet line), and accumulative thermal conductivity $\kappa_a$ (blue line) for 2 typical realizations in a 4×4×4 super cell. A total time of 400ps is plotted in the log-scale. Insets in (b) and (e) show the time dependence of the mean value of the normalized HCACF (purple line). The green and black arrows pinpoint the cut-off time estimated by first avalanche and first dip, respectively. The black line draws the zero-axis for reference.

FIG. 2. (Color online) Thermal conductivity of the crystalline Silicon (a) and Germanium (b) at 1000K versus super cell size (N×N×N unit cells) for different methods with first avalanche. The symbols square, dot, and triangle denote calculation results based on direct integration and double-exponential-fitting without/with $Y_0$, respectively.

FIG. 3. (Color online) Raw data (red line) and the corresponding fitted curve according to double exponential fitting (blue line) of the normalized heat current autocorrelation function before the cut-off time (denoted by green arrow) in a 4×4×4 super cell for the same two realizations shown in Figure 1. The insets show the long time region near the cut-off time. The black line draws the zero-axis for reference.

TABLE I. EMD simulation results for thermal conductivity of crystalline Silicon at 1000K. Experimental value is 31.0 W/mK from Ref. [23]. A cubic super cell of N×N×N unit cells is used, and the unit cell length of Si is 0.543nm. $\kappa_D$ denotes the calculation results using direct integration. $\kappa_C$ and $\kappa_F$ denote the calculation results using double exponential fitting without and with $Y_0$, respectively.

| N | Number of atoms | $\kappa_D$ (W/mK) | $\kappa_C$ (W/mK) | $\kappa_F$ (W/mK) |
|---|---|---|---|---|
| 4 | 512 | 17.78±2.61 | 10.27±0.63 | 17.59±2.66 |
| 6 | 1728 | 22.36±3.62 | 12.69±1.10 | 22.39±3.62 |
| 8 | 4096 | 26.41±4.02 | 14.47±1.27 | 26.42±4.01 |
| 10 | 8000 | 30.72±3.24 | 15.90±1.49 | 30.74±3.23 |
| 12 | 13824 | 31.48±2.86 | 16.31±1.57 | 31.50±2.85 |

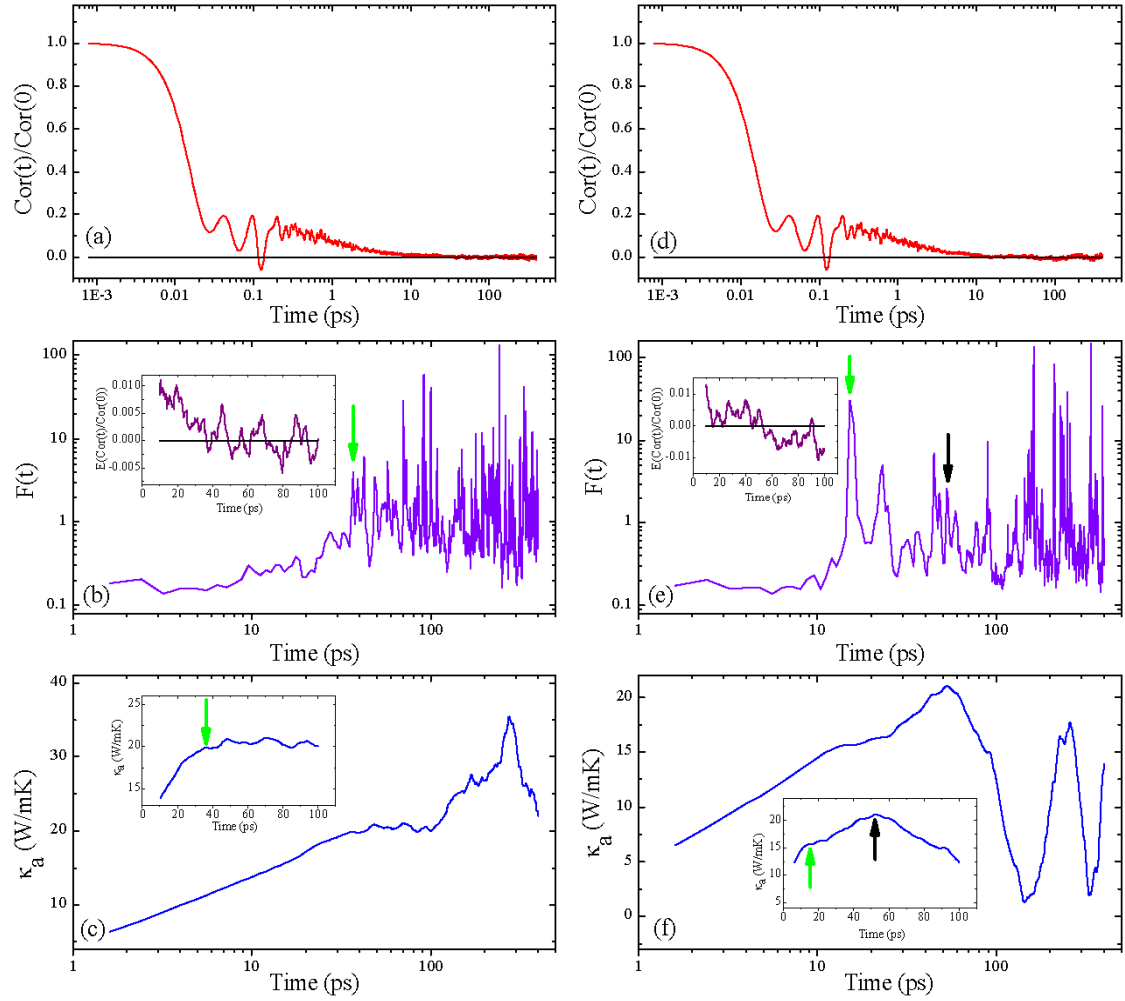

FIG. 1. (Color online) Time dependence of normalized heat current autocorrelation function Cor(t)/Cor(0) (red line), relative fluctuation of Cor(t) (violet line), and accumulative thermal conductivity $\kappa_a$ (blue line) for 2 typical realizations in a 4×4×4 super cell. A total time of 400ps is plotted in the log-scale. Insets in (b) and (e) show the time dependence of the mean value of the normalized HCACF (purple line). The green and black arrows pinpoint the cut-off time estimated by first avalanche and first dip, respectively. The black line draws the zero-axis for reference.

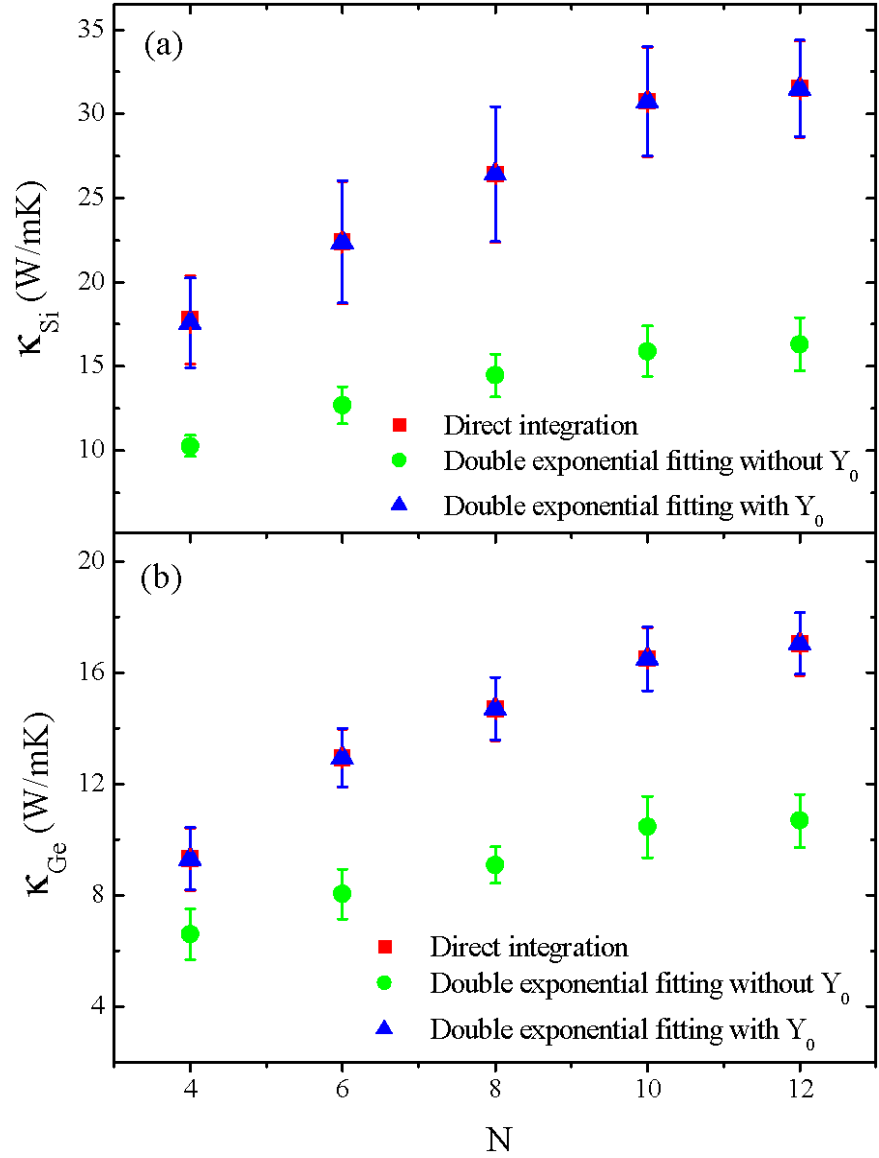

FIG. 2. (Color online) Thermal conductivity of the crystalline Silicon (a) and Germanium (b) at 1000K versus super cell size (N×N×N unit cells) for different methods with first avalanche. The symbols square, dot, and triangle denote calculation results based on direct integration and double-exponential-fitting without/with $Y_0$, respectively.

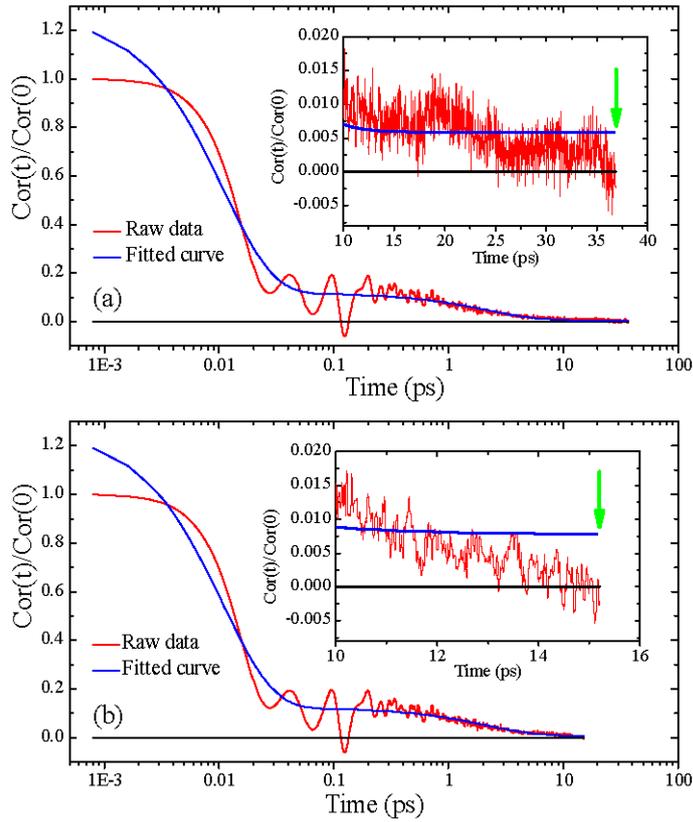

FIG. 3. (Color online) Raw data (red line) and the corresponding fitted curve according to double exponential fitting (blue line) of the normalized heat current autocorrelation function before the cut-off time (denoted by green arrow) in a 4×4×4 super cell for the same two realizations shown in Figure 1. The insets show the long time region near the cut-off time. The black line draws the zero-axis for reference.